\documentclass[pra,preprint,showkeys]{revtex4}
\usepackage{amsmath}
\usepackage{amssymb}
\usepackage{revsymb}
\usepackage{mathrsfs}
\begin{document}
\title{Exact solution of the Landau-Lifshitz equation in a plane wave}
\author{A. \surname{Di Piazza}\email{dipiazza@mpi-hd.mpg.de}}
\affiliation{Max-Planck-Institut f\"ur Kernphysik, Saupfercheckweg 1, D-69117 Heidelberg, Germany}

\begin{abstract}
The Landau-Lifshitz \cite{Landau_b_2_1975} form of the Lorentz-Abraham-Dirac equation in the presence of a plane wave of arbitrary shape and polarization is solved exactly and in closed form. The explicit solution is presented in the particular, paradigmatic cases of a constant crossed field and of a monochromatic wave with circular and with linear polarization.

\vspace{1cm}


\end{abstract}

\maketitle

An accelerated electric charge emits electromagnetic radiation. The Lorentz-Abraham-Dirac (LAD) equation is the equation of motion of an electric charge in the presence of an external field when the back-reaction of the radiated field onto the motion of the charge is taken into account consistently and in a covariant way \cite{Landau_b_2_1975}. This equation was originally derived in the nonrelativistic limit by Lorentz and then generalized to the relativistic case by Abraham and Dirac \cite{Abraham_1902,Abraham_1904,Dirac_1938}. It is well known that the LAD equation contains the time derivative of the acceleration of the charge (an electron, for definites) and this causes serious difficulties related to the appearance of ``runaway'' solutions, i. e. of solutions that show an exponential increase of the electron acceleration even in the absence of external fields (see \cite{Klepikov_1985,Rohrlich_b_1990,Rohrlich_2000,Spohn_b_2004} for a review). Dirac proposed to overcome this difficulty by postulating that the physical solutions of the LAD equation must satisfy the asymptotic condition of zero-acceleration at late times. However, the validity of this condition has been tested only in a few cases and, in general, it is not excluded that many solutions may fulfill it (see \cite{Spohn_2000} and References therein). A different way to avoid the occurrence of runaway solutions has been proposed in \cite{Landau_b_2_1975} where a perturbative reduction of order is performed on the LAD equation. The resulting equation which we will indicate as the Landau-Lifshitz (LL) equation, does not contain the time derivative of the electron acceleration and is therefore not plagued by the existence of runaway solutions. Although the general validity of the LAD equation, the LL equation has been derived in \cite{Landau_b_2_1975} only in the case of an external electromagnetic field and we limit to this case here. Other quantitative aspects of the LL equation will be discussed below. We point out here that its importance in the framework of classical electrodynamics has been reinforced by the rigorous proof that the physical solutions of the LAD equation are on the critical manifold and are governed there exactly by the LL equation \cite{Spohn_2000,Spohn_b_2004} (see also \cite{Rohrlich_2001}). Both the LAD and the LL equations are classical equations and they are valid if quantum effects are negligible. In general, this is the case when the recoil of the radiated photons onto the electron can be neglected \cite{Nikishov_1964_a,Nikishov_1964_b,Nikishov_1965}. In \cite{Landau_b_2_1975} it is shown that for a large range of electron and external field parameters the LL equation can be consistently applied to different problems where the radiation back-reaction force (or the so-called ``self-force'' \cite{Rohrlich_2000}) is even comparable with the Lorentz force (see also \cite{Shen_1970,Shen_1972}). 

Intense laser beams represent a unique experimental tool to investigate classical and quantum electrodynamics in the presence of strong fields (see the recent reviews \cite{Salamin_2006,Mourou_2006} and the References therein). A record intensity of $7\times 10^{21}\;\text{W/cm$^2$}$ has already been obtained in laboratory by focusing a 45-TW Ti:Sa laser beam onto one wavelength \cite{Bahk_2004}. Numerous Petawatt laser systems in principle capable of reaching an intensity of about $10^{23}\;\text{W/cm$^2$}$, are under construction in different laboratories \cite{Norby_2005}. Also, in the near future the Extreme Light Infrastructure (ELI) is expected to reach unprecedented intensities of about $10^{25}\text{-}10^{26}\;\text{W/cm$^2$}$ \cite{ELI}. Finally, theoretical schemes have been proposed to obtain the ultimate Schwinger intensity of approximately $10^{29}\;\text{W/cm$^2$}$ corresponding to an electric (magnetic) field amplitude of $E_{cr}\equiv m^2c^3/\hbar e=1.3\times 10^{16}\;\text{V/cm}$ ($B_{cr}\equiv m^2c^3/\hbar e=4.4\times 10^{13}\;\text{G}$), at which electron-positron pairs are spontaneously created from vacuum (here $-e<0$ and $m$ are the electron charge and mass, respectively) \cite{Tajima_2002,Gordienko_2005}. In the presence of such intense electromagnetic fields the dynamics of the electron is strongly influenced by its own radiation and testing the validity of the LL equation experimentally will become possible. 

In this Letter we present the exact analytical solution of the LL equation in closed form in the presence of a general plane wave of arbitrary shape and polarization. To the best of our knowledge, no such solution has been found so far. The LL equation in the presence of a constant and uniform electromagnetic field has been solved exactly \cite{Shen_1970,Shen_1972,Herrera_1973,Heintzmann_1973,Herrera_1974,Ares_De_Parga_1999} and numerous applications to astrophysics and to synchrotron radiation have been discussed. In \cite{Schrufer_t_1973} a general method of solving the LL equation in a plane wave is presented; however, as a result, the solution is given in implicit form and the solution of one integro-differential equation is required. An application of this method can be found in \cite{Grewing_1973} in the particular case of a monochromatic, linearly polarized plane wave but, as we will see, the solution found is incorrect (see also \cite{Heintzmann_1972}). Numerical approaches to the LL equation in a plane wave can be found in the literature, for studying the effects of the radiation back-reaction on the electron trajectory \cite{Keitel_1998} and on the frequency spectrum of the radiation emitted by the electron \cite{Keitel_1998,Bulanov_2004,Koga_2004,Koga_2005}. In particular, in \cite{Koga_2005} the so-called radiation-dominated regime is investigated where the self-force is comparable or even larger than the Lorentz force. Finally, exact numerical and perturbative analytical approaches to the LAD equation in a plane wave are presented in \cite{Gora_1951,Jackson_1984,Hartemann_1996,Hartemann_1998,Hartemann_2005}. 

In the following, natural units with $\hbar=c=1$ are used and the fine-structure constant is given by $\alpha\equiv e^2/4\pi\approx 1/137$. Finally, the Minkowski metric tensor is $\eta^{\mu\nu}=\text{diag}(+1,-1,-1,-1)$ and the four-dimensional product of two generic four-vectors $a^{\mu}$ and $b^{\mu}$ is indicated as $(ab)$, i. e. $(ab)\equiv a^{\mu}b_{\mu}$. In the particular case that $a^{\mu}=b^{\mu}$ we employ the short notation $a^2\equiv (aa)$.

The relativistic covariant form of the LAD equation for an electron in a general electromagnetic field $F^{\mu\nu}(x)$ reads \cite{Landau_b_2_1975}
\begin{equation}
\label{LAD_eq}
m\frac{d u^{\mu}}{ds}=-eF^{\mu\nu}(x)u_{\nu}+\frac{2}{3}\alpha\left(\frac{d^2 u^{\mu}}{ds^2}+\frac{d u^{\nu}}{ds}\frac{d u_{\nu}}{ds}u^{\mu}\right).
\end{equation}
In this equation, $x^{\mu}$ and $u^{\mu}\equiv dx^{\mu}/ds$ are the spacetime coordinates and the four-velocity of the electron, respectively, and $s$ is its proper time. The terms proportional to $\alpha$ in Eq. (\ref{LAD_eq}) represent the self-force due to the radiation back-reaction. The approach used in \cite{Landau_b_2_1975} to reduce Eq. (\ref{LAD_eq}) to a second-order differential equation in the electron coordinates $x^{\mu}$ consists in assuming that in the rest frame of the electron the self-force is much smaller than the Lorentz force and in substituting the derivatives of the four-velocity in the self-force with their zero-order expression $-eF^{\mu\nu}(x)u_{\nu}/m$. The resulting LL equation is (by omitting, for simplicity, the dependence of $F^{\mu\nu}(x)$ on the electron coordinates)
\begin{equation}
\label{LL_eq}
\begin{split}
m\frac{d u^{\mu}}{ds}&=-eF^{\mu\nu}u_{\nu}-\frac{2}{3}\alpha\left[\frac{e}{m}(\partial_{\alpha}F^{\mu\nu})u^{\alpha}u_{\nu}\right.\\
&\left.\qquad+\frac{e^2}{m^2}F^{\mu\nu}F_{\alpha\nu}u^{\alpha}-\frac{e^2}{m^2}(F^{\alpha\nu}u_{\nu})(F_{\alpha\lambda}u^{\lambda})u^{\mu}\right].
\end{split}
\end{equation}
As it results from the derivation given in \cite{Landau_b_2_1975}, this equation is valid when the typical frequency of the external electromagnetic field measured in the rest system of the electron is much smaller than the electron mass and when its typical strength measured in the rest system of the electron is much smaller than $E_{cr}/\alpha\approx 137\, E_{cr}$ and $B_{cr}/\alpha\approx 137\, B_{cr}$ (we will state this more precisely below). However, there is a further condition on the external field strength to be fulfilled in the framework of classical electrodynamics that is more stringent than the above one. In fact, as it is shown in \cite{Nikishov_1964_a,Nikishov_1964_b,Nikishov_1965}, quantum effects become important already when the typical strength of the electromagnetic field measured in the rest system of the electron is of order of the critical fields $E_{cr}$ and $B_{cr}$. 

The purpose of this Letter is to find the exact analytical solution of Eq. (\ref{LL_eq}) in closed form for the case when the external field is represented by a general plane wave, i. e. a field that depends on the spacetime coordinates $x^{\mu}$ only through the combination $\phi\equiv (nx)$ where $n^{\mu}\equiv(1,\mathbf{n})$ satisfies the condition $n^2=0$, with $\mathbf{n}$ being the wave propagation direction. We indicate as $A^{\mu}(\phi)$ the four-vector potential corresponding to the electromagnetic field $F^{\mu\nu}(\phi)\equiv\partial^{\mu}A^{\nu}(\phi)-\partial^{\nu}A^{\mu}(\phi)$ and we decide to work in the Lorentz gauge with $\partial_{\mu}A^{\mu}(\phi)=0$. The general expression of the four-vector potential of a plane wave with arbitrary polarization is given by 
\begin{equation}
\label{A}
A^{\mu}(\phi)=a_1^{\mu}\psi_1(\phi)+a_2^{\mu}\psi_2(\phi).
\end{equation}
In this expression $a_j^{\mu}$ with $j\in \{1,2\}$ are two constant four-vectors such that $(na_j)=0$ and $(a_ja_{j'})=a_j^2\delta_{jj'}$, with $\delta_{jj'}$ being the Kronecker delta-function. Also, $\psi_j(\phi)$ are two arbitrary scalar functions of $\phi$. By introducing the two constant quantities
\begin{equation}
f_j^{\mu\nu}\equiv n^{\mu}a_j^{\nu}-n^{\nu}a_j^{\mu},
\end{equation}
the electromagnetic field $F^{\mu\nu}(\phi)$ of the wave can be written as
\begin{equation}
\label{F_mu_nu}
F^{\mu\nu}(\phi)=f_1^{\mu\nu}\psi'_1(\phi)+f_2^{\mu\nu}\psi'_2(\phi),
\end{equation}
where the prime indicates the derivative with respect to $\phi$ (note that since $\phi$ is a dimensional variable, the quantities $f_j^{\mu\nu}$ do not have the dimensions of an electromagnetic field). The constant quantities $f_j^{\mu\nu}$ satisfy the relations $n_{\mu}f_j^{\mu\nu}=0$ and $f_j^{\mu\nu}f_{j',\lambda\nu}=a_j^2\delta_{jj'}n^\mu n_{\lambda}$. It is worth noting that $F^{0\nu}(\phi)F_{\nu 0}(\phi)\equiv E^2(\phi)$ with $E^2(\phi)$ being the square of the wave electric field. Since on the other hand $F^{0\nu}(\phi)F_{\nu 0}(\phi)=-[a_1^2(\psi'_1(\phi))^2+a_2^2(\psi'_2(\phi))^2]$, then it must be $a_j^2<0$ with $j\in \{1,2\}$.

The key idea to solve Eq. (\ref{LL_eq}) is to use the quantity $\phi$ as the dependent variable and to express all the quantities as functions of $\phi$. It is clear that in this method the derivative $ds/d\phi$ of the proper time with respect to $\phi$ plays an important role:
\begin{equation}
\frac{ds}{d\phi}=\left(\frac{d\phi}{ds}\right)^{-1}=\frac{1}{(nu(\phi))}.
\end{equation}
By multiplying the LL equation (\ref{LL_eq}) by $n^{\mu}$ and by using the orthogonality properties of the quantities $f_j^{\mu\nu}$ it can be shown that the proper time $s$ satisfies the differential equation
\begin{equation}
m\frac{d^2s}{d\phi^2}=\frac{2}{3}\alpha[\xi_1^2(\psi'_1(\phi))^2+\xi_2^2(\psi'_2(\phi))^2],
\end{equation}
where we have introduced the two invariant field strength parameters $\xi_j^2\equiv -e^2a_j^2/m^2$. By integrating this equation with respect to $\phi$ once, we obtain the general expression of the quantity $\rho(\phi)\equiv (nu(\phi))$ as a function of $\phi$:
\begin{equation}
\rho(\phi)=\frac{\rho_0}{h(\phi)}.
\end{equation}
In this expression we have introduced the parameter $\rho_0\equiv (nu_0)$ relative to the four-velocity $u_0^{\mu}\equiv u^{\mu}(\phi_0)$ at the initial value $\phi_0$ and the function
\begin{equation}
h(\phi)\equiv 1+\frac{2}{3}\alpha\frac{\rho_0}{m}\int_{\phi_0}^{\phi}d\varphi[\xi_1^2(\psi'_1(\varphi))^2+\xi_2^2(\psi'_2(\varphi))^2].
\end{equation}
The initial four-velocity $u_0^{\mu}$ is subject to the physical, invariant condition $(u_0u_0)=1$ and the LL equation (\ref{LL_eq}) guarantees that this condition is fulfilled at any $\phi$ by the four-velocity $u^{\mu}(\phi)$. By defining the ``reduced'' four-velocity $\tilde{u}^{\mu}(\phi)\equiv h(\phi)u^{\mu}(\phi)$ it can easily be shown that the LL equation (\ref{LL_eq}) in the presence of the plane wave field (\ref{F_mu_nu}) becomes
\begin{equation}
\begin{split}
m\frac{d\tilde{u}^{\mu}}{d\phi}&=-\frac{1}{\rho_0}\left[h(\phi)eF^{\mu\nu}(\phi)+\frac{2}{3}\alpha\frac{\rho_0}{m}eF^{'\mu\nu}(\phi)\right]\tilde{u}_{\nu}\\
&\quad+\frac{2}{3}\alpha h(\phi)[\xi_1^2(\psi'_1(\phi))^2+\xi_2^2(\psi'_2(\phi))^2]n^{\mu}.
\end{split}
\end{equation}
This equation has the general structure of a linear system of first-order, ordinary differential equations with coefficients depending on the integration variable and it can be formally solved by iteration. In general, this procedure leads to a solution expressed in the form of an infinite series. However, due to the particular structure of a general plane wave and, in particular, to the orthogonality properties of the \emph{constant} quantities $f_j^{\mu\nu}$, it can be shown that only two iterations are required, the higher order terms being identically zero. The procedure is rather lengthy but straightforward and we present here only the final exact solution
\begin{equation}
\label{Sol}
\begin{split}
u^{\mu}(\phi)&=\frac{1}{h(\phi)}\left\{u_0^{\mu}+\frac{1}{2\rho_0}[h^2(\phi)-1]n^{\mu}-\frac{1}{\rho_0}\left[\mathcal{I}_1(\phi)\frac{ef_1^{\mu\nu}}{m}+\mathcal{I}_2(\phi)\frac{ef_2^{\mu\nu}}{m}\right]u_{0,\nu}\right.\\
&\qquad\qquad+\frac{1}{2\rho_0}[\xi_1^2\mathcal{I}^2_1(\phi)+\xi_2^2\mathcal{I}^2_2(\phi)]n^{\mu}\Bigg\},
\end{split}
\end{equation}
where
\begin{equation}
\begin{split}
\mathcal{I}_j(\phi)&\equiv\int_{\phi_0}^{\phi}d\varphi\left[h(\varphi)\psi'_j(\varphi)+\frac{2}{3}\alpha\frac{\rho_0}{m}\psi''_j(\varphi)\right]\\
&=\psi_j(\phi)h(\phi)-\psi_j(\phi_0)+\frac{2}{3}\alpha\frac{\rho_0}{m}[\psi'_j(\phi)-\psi'_j(\phi_0)]\\
&\quad-\frac{2}{3}\alpha\frac{\rho_0}{m}\int_{\phi_0}^{\phi}d\varphi\psi_j(\varphi)[\xi_1^2(\psi'_1(\varphi))^2+\xi_2^2(\psi'_2(\varphi))^2].
\end{split}
\end{equation}
Our solution satisfies the constraint $(u(\phi)u(\phi))=1$ identically and in the absence of radiation back-reaction (formally in the limit $\alpha\to 0$), it reduces to the standard solution of the Lorentz equation in a plane wave \cite{Landau_b_2_1975}. The solution presented in \cite{Grewing_1973} for the particular case of a monochromatic, linearly polarized wave does not contain the term proportional to $h^2(\phi)-1$ (see Eq. (\ref{Sol})) which is already crucial to fulfilling the physical constraint $(u(\phi)u(\phi))=1$. It is worth noting that our general solution is relatively compact and that the integrations involved are in principle no more difficult than those already present in the textbook solutions in the absence of radiation back-reaction. Once the four-velocity is obtained, the electron spacetime coordinates $x^{\mu}(\phi)$ as functions of $\phi$ are given by
\begin{equation}
x^{\mu}(\phi)=x^{\mu}_0+\frac{1}{\rho_0}\int_{\phi_0}^{\phi}d\varphi h(\varphi)u^{\mu}(\varphi),
\end{equation}
with $x^{\mu}_0\equiv x^{\mu}(\phi_0)$.

Below, we discuss the explicit expressions of the functions $h(\phi)$ and $\mathcal{I}_j(\phi)$ in the paradigmatic cases of a constant crossed field (CCF), of a monochromatic, circularly polarized wave (MCPW) and of a monochromatic, linearly polarized wave (MLPW). For a CCF the four-vector potential can be chosen as $A^{\mu}(\phi)=(0,\mathbf{A}(\phi))$ where $\mathbf{A}(\phi)=-E\phi\mathbf{e}$, with $\mathbf{e}$ being a unit vector perpendicular to $\mathbf{n}$ and $E$ being the electric field amplitude. A convenient choice of the functions $\psi_j(\phi)$ is: $\psi_1(\phi)=E\phi/m$, then $a_1^{\mu}=(0,-m\mathbf{e})$ and $\psi_2(\phi)=0$. In this case we obtain $\xi_1^2=e^2$ and that
\begin{align}
h(\phi)&=1+\frac{2}{3}\alpha\rho_0\frac{E^2}{E_{cr}^2}m(\phi-\phi_0),\\
\mathcal{I}_1(\phi)&=\frac{E}{m}\left[\phi-\phi_0+\frac{1}{3}\alpha\rho_0\frac{E^2}{E_{cr}^2}m(\phi-\phi_0)^2\right],
\end{align}
while $\mathcal{I}_2(\phi)=0$. We note that it is not important how we choose the functions $\psi_j(\phi)$ and the four-vectors $a_j^{\mu}$. In fact, in the final solution Eq. (\ref{Sol}) these quantities always appear in the combinations $\psi_j(\phi)a_j^{\mu}$. 

In the case of a MCPW with frequency $\omega_0$ and amplitude $E$ we have $\psi_1(\phi)=\cos(\omega_0\phi)$, $\psi_2(\phi)=\sin(\omega_0\phi)$ and $\xi_1^2=\xi_2^2=(eE/m\omega_0)^2\equiv\xi^2$. The results for the functions $h(\phi)$ and $\mathcal{I}_j(\phi)$ are
\begin{align}
h(\phi)&=1+\frac{2}{3}\alpha\eta_0\xi^2\omega_0(\phi-\phi_0),\\
\mathcal{I}_1(\phi)&=\cos(\omega_0\phi) h(\phi)-\cos(\omega_0\phi_0)-\frac{2}{3}\alpha\eta_0(1+\xi^2)(\sin(\omega_0\phi)-\sin(\omega_0\phi_0)),\\
\mathcal{I}_2(\phi)&=\sin(\omega_0\phi)h(\phi)-\sin(\omega_0\phi_0)+\frac{2}{3}\alpha\eta_0(1+\xi^2)(\cos(\omega_0\phi)-\cos(\omega_0\phi_0)),
\end{align}
where we have introduced the invariant parameter $\eta_0\equiv\omega_0\rho_0/m=(k_0u_0)/m$ with $k_0^{\mu}\equiv\omega_0n^{\mu}$ being the four-dimensional wave-vector of the wave. 

Finally, in the case of a MLPW with frequency $\omega_0$ and amplitude $E$ we have $\psi_1(\phi)=\cos(\omega_0\phi)$, $\psi_2(\phi)=0$ and $\xi_1^2=\xi^2$. The results for the functions $h(\phi)$ and $\mathcal{I}_j(\phi)$ then are
\begin{align}
h(\phi)&=1+\frac{1}{3}\alpha\eta_0\xi^2[\omega_0(\phi-\phi_0)-\sin(\omega_0\phi)\cos(\omega_0\phi)+\sin(\omega_0\phi_0)\cos(\omega_0\phi_0)],\\
\begin{split}
\mathcal{I}_1(\phi)&=\cos(\omega_0\phi) h(\phi)-\cos(\omega_0\phi_0)\\
&-\frac{2}{3}\alpha\eta_0\left[1+\frac{\xi^2}{3}(\sin^2(\omega_0\phi)+\sin(\omega_0\phi)\sin(\omega_0\phi_0)+\sin^2(\omega_0\phi_0))\right]\\
&\qquad\times(\sin(\omega_0\phi)-\sin(\omega_0\phi_0)),
\end{split}
\end{align}
while $\mathcal{I}_2(\phi)=0$. In both cases of a circularly and linearly polarized monochromatic wave we see that the important parameters of the problem are $\eta_0$ and $\xi$. $\eta_0$ is the wave frequency measured in the rest frame of the electron in unit of the electron mass, whereas $\xi$ is the work carried out by the wave on the electron in a distance of order of one wavelength in unit of the electron mass. In the relevant cases where the electron in the wave is highly relativistic ($\xi\gg 1$) our solution shows that the radiation back-reaction effects scale essentially with the parameter $R\equiv\alpha\eta_0\xi^2$ (see also \cite{Koga_2005}). From the relations $\eta_0=(k_0u_0)/m$ and $\xi=eE/m\omega_0$ it is clear that the quantity $\eta_0\xi$ is exactly the parameter $\chi$ already introduced below Eq. (\ref{LL_eq}), i. e. the wave electric field amplitude in the rest frame of the electron measured in units of the critical field. Since the LL equation is valid when $\eta_0,\chi\ll 1$, we see that at moderate electron energies and sufficiently intense waves the classical treatment is suitable for investigating situations in which the effects of the self-force are even of the same order of the Lorentz force and cannot be treated perturbatively. For example, if we consider an optical laser field with $\omega_0=1\;\text{eV}$ and with an intensity of $10^{24}\;\text{W/cm$^2$}$, then $\xi\approx 10^3$, and for electrons propagating in the opposite direction of the laser field with an energy of about $20\;\text{MeV}$ we have $\eta_0\approx 2\times 10^{-4}$, $\chi\approx 0.2$ and $R\approx 1.3$. In this regard, the situation is complementary to that in a constant and uniform magnetic field where the optimal region of parameters for a classical treatment to be valid in the radiation-reaction dominating regime is for relatively weak magnetic fields and highly relativistic electrons \cite{Shen_1970,Shen_1972}.

In conclusion, we have found the exact analytical solution of the LL equation in the presence of a plane electromagnetic wave of arbitrary shape and polarization. We have given the explicit expression of the solution in some paradigmatic cases like those of a constant crossed field as well as of a monochromatic field with circular and with linear polarization. Our solution can be applied to numerous physical situations in which the electron is subject to violent accelerations in the presence of strong laser fields. For example, it can be employed as a means to investigate the generation of monochromatic high-energy electron beams by laser fields, the electromagnetic emission by electrons in intense laser fields and in the wide field of strong laser-plasma interaction.

The author gratefully acknowledges valuable discussions with K. Z. Hatsagortsyan. 
%
%*****************************************************************
%

\end{document}